\documentclass[doublecol]{epl2}
\usepackage{graphicx}
\usepackage{subfigure}
\usepackage{cite}
\usepackage{amssymb}
\usepackage{newlfont}

\title{Measuring robustness of community structure in complex networks}
\shorttitle{Measuring robustness of community structure in complex networks}

\author{Hui-Jia Li\inst{1}\thanks{Corresponding authors: \email{Hjli@amss.ac.cn}}, Hao Wang\inst{1}\thanks{Corresponding authors: \email{holy.wong@connect.polyu.hk}}, Luonan Chen\inst{2}}

\shortauthor{Hui-Jia Li \etal}

\institute{
  \inst{1} School of Management Science and Engineering, Central University of Finance and Economics, Beijing 100080, China.\\
  \inst{2} Key Laboratory of Systems Biology, Shanghai Institutes for Biological Sciences, Chinese Academy of Sciences, Shanghai 200233, China.}
\abstract{The theory of community structure is a powerful tool for real networks, which can simplify their topological and functional analysis considerably. However, since community detection methods have random factors and real social networks obtained from complex systems always contain error edges, evaluating the robustness of community structure is an urgent and important task.
In this letter, we employ the critical threshold of resolution parameter in Hamiltonian function, $\gamma_C$, to measure the robustness of a network. According to spectral theory, a rigorous proof shows that the index we proposed is inversely proportional to robustness of community structure.
Furthermore, by utilizing the co-evolution model, we provides a new efficient method for computing the value of $\gamma_C$.
The research can be applied to broad clustering problems in network analysis and data mining due to its solid mathematical basis and experimental effects.}

\begin{document}

\maketitle

\section{ 1. Introduction}
Community structure detection \cite{Newman}-\cite{X} is a hotspot of
social network studies. It has attracted much
attention from various scientific fields. Generally, community refers
to a group of nodes in the network that are more densely connected
internally than the rest of the network. A well known exploration for this problem is the concept of modularity, which
is proposed by Newman et al.\cite{Newman}\cite{Girvan} to quantify a network's partition. Optimizing modularity is effective for community
structure detection and has been widely used in many real networks\cite{X}.
However, as pointed out by Fortunato et al.\cite{Fortunato},
modularity is restricted by the resolution limit problem which is concerned
about the reliability of the communities detected through the
optimization methods. Complementary to the modularity concept, many efforts are devoted to understanding the
properties of dynamic processes taking place in the underlying
networks. Specifically, researchers have begun to investigate the
correlation between community structure and
dynamic systems, such as synchronization\cite{Arenas1} and random walk process\cite{Weinan}.

\begin{figure}
\centering
  \subfigure[]{
    \label{fig:subfig:1a} 
    \setcounter{subfigure}{1}
    \includegraphics[width=7.3cm,height=3.8cm]{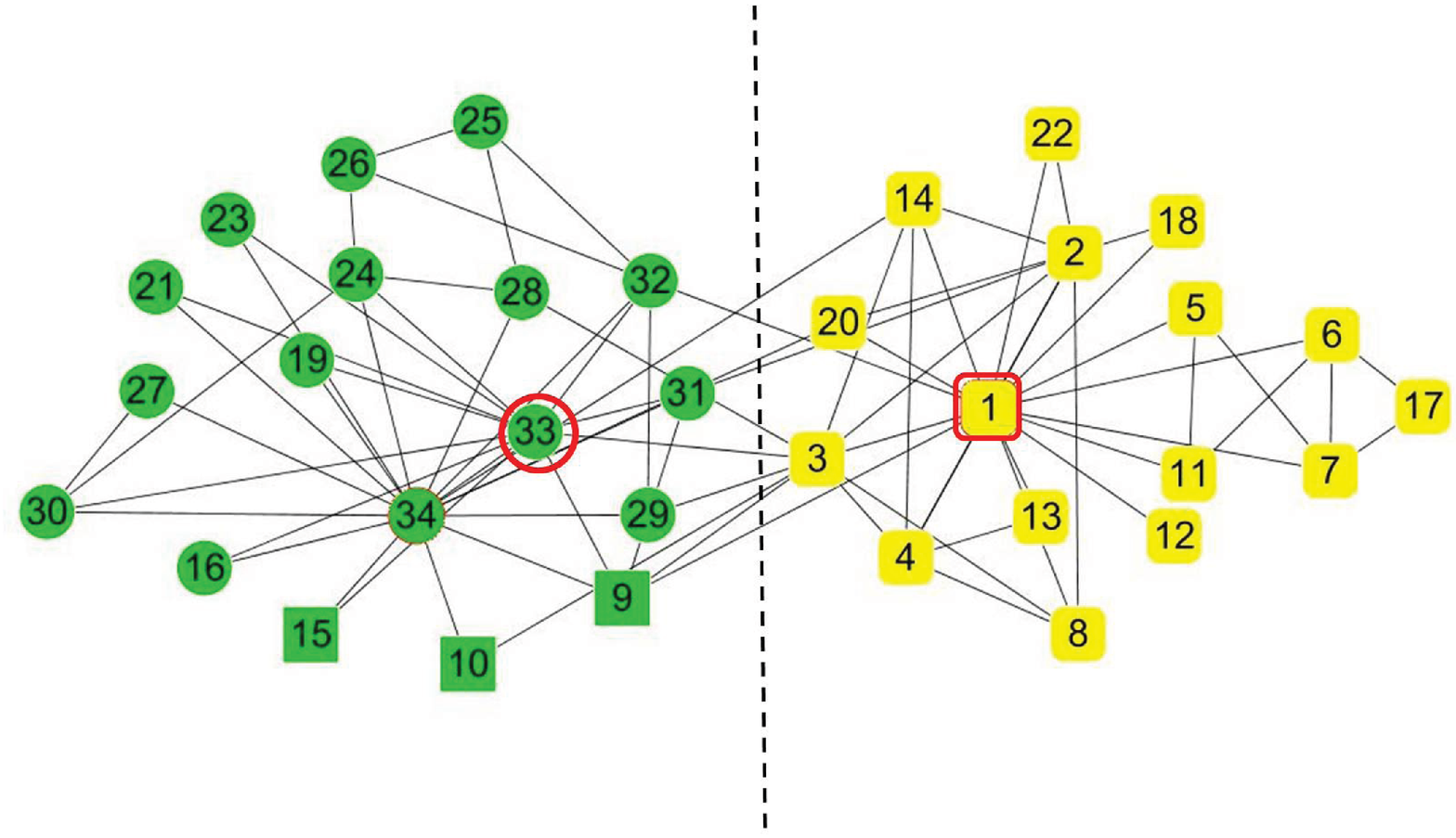}}
  \subfigure[]{
    \label{fig:subfig:1b} 
    \setcounter{subfigure}{2}
    \includegraphics[width=6.5cm,height=1.5cm]{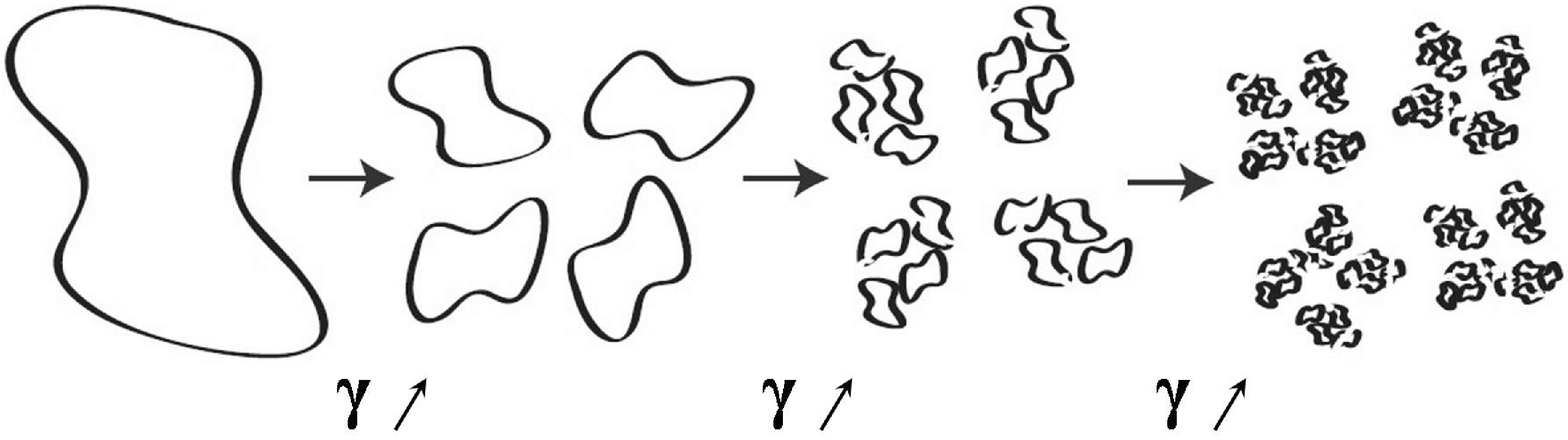}}
  \subfigure[]{
    \label{fig:subfig:1c} 
    \setcounter{subfigure}{3}
    \includegraphics[width=8cm,height=4.3cm]{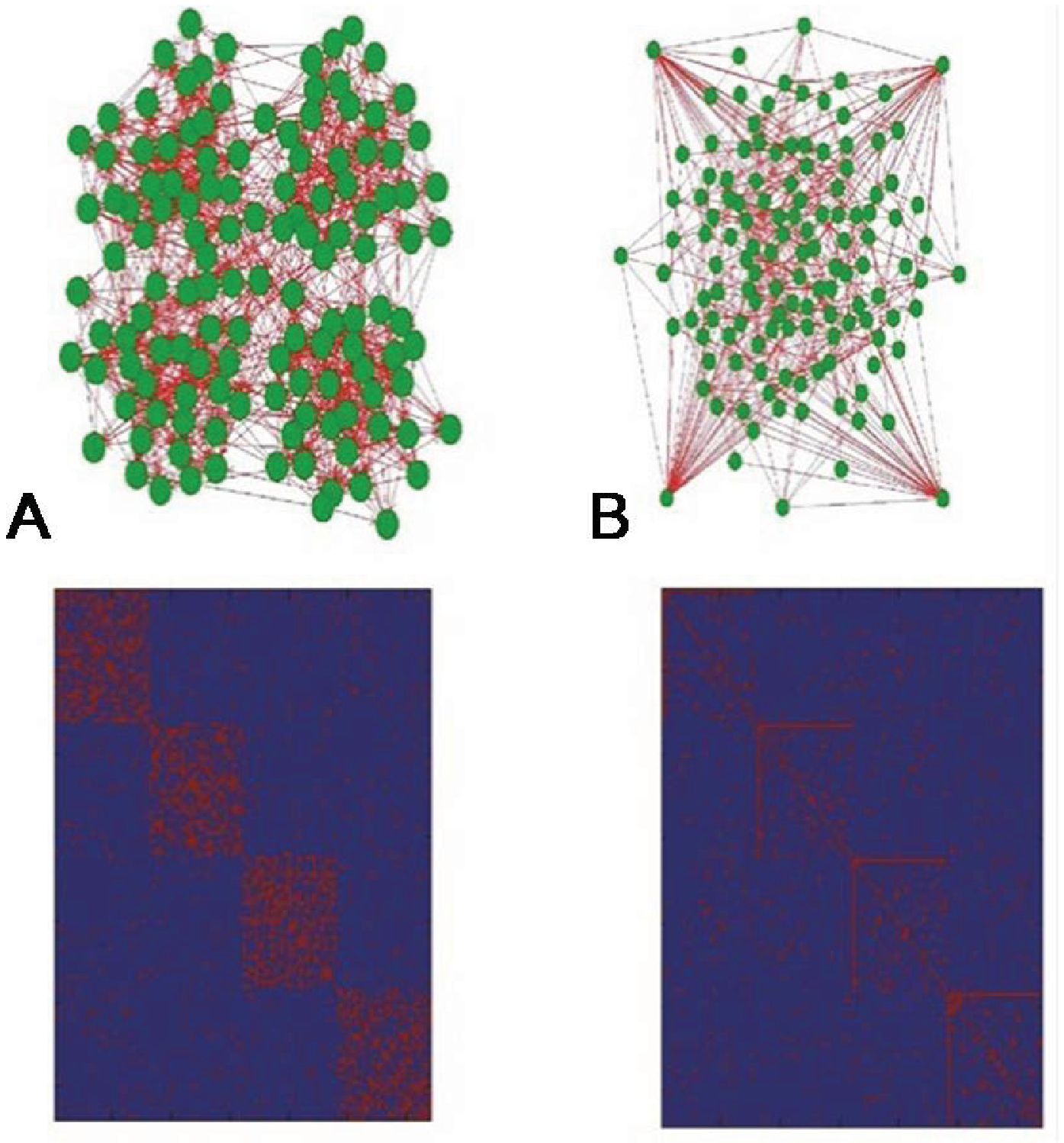}}
\caption{(a) The community structure of the karate club network detected
by Wayne Zachary; (b) A network can be partitioned into more and more communities when $\gamma$ increases;(c) A strong 4-community structure in network $A$ and a weak 4-community structure in network $B$.} \label{Fig:1}
\end{figure}

In the real-world, network topology
changes over time. The analysis of community structure in evolving networks has been
regarded as a ``Holy Grail" of network scientists.
A famous example is the karate club network constructed by Wayne Zachary in 1970s\cite{Zachary}.
During the course of his study, a dispute arose between the club's administrator and principal karate teacher over whether to raise club fees, and the club eventually split into two smaller
clubs, centered around the administrator(node 1) and the teacher(node 33), as shown in Fig.\ref{fig:subfig:1a}.
It can be assumed that, the relationships between members in karate club at the very beginning are not robust and small perturbation
may cause the complete change of the topology.
Why the initial un-tight relationship evolves into or out of community structure is a very interesting question\cite{Karrer},
since community structure has a great impact on human organizational structure,
rumor and epidemic spreading, network attack effect and congestion control.

Given a network, it is meaningless to detect the community when the community structure is un-robust: if a small change in the network, for example, an edge added here or there, can completely change the outcome(significance or stability)
of community structure, then, we argue that,
the network is un-robust and the result could not be trustworthy.
In this letter we focus on this imperative task and prove the critical threshold of resolution parameter in Hamiltonian function, $\gamma_C$, can measure the robustness of a network.

For any
given network, robustness information can be derived from $\gamma_C$ directly and conveniently without using particular partition algorithms.
A rigorous proof is then used to show the index we proposed is inversely proportional to robustness of community structure theoretically based on spectral theory. Furthermore, to calculate the value of $\gamma_C$, a new efficient method is provided based on co-evolution model.
The method can be applied to broad clustering problems in network analysis and data mining due to its solid mathematical basis and efficiency.

\section{ 2. Potts model and the ciritcal resolution parameter $\gamma$}
Potts model\cite{Reichardt} is a powerful thermodynamic method, which has been widely applied to uncover community structure in networks.
We use the multiresolution
Potts method to carry out the study. Given a network $G$ and corresponding adjacency matrix
$A=\{A_{ij}\}$, community structure can be determined by minimizing the
infinite $N$-state Hamiltonian function:

\begin{equation} \label{eq:1}
H(\gamma)=\sum_{i\neq j}J_{ij}(\gamma)\delta_{C_i,C_j}\\
=\sum_{i\neq j}(A_{ij}-\gamma P_{ij})\delta_{C_i,C_j},
\end{equation}
where $C_i$ represents the community (state) that node (spin) $i$ belongs, $\gamma$ is the resolution parameter, $P_{ij}$
represents the expected number of edges between nodes $i$ and $j$ in the null model, and $J(\gamma)$ is the coupling
matrix with entries $J_{ij}$ represents the interaction strength between node $i$ and $j$.

In Potts model, the resolution parameter $\gamma$ is an important indicator of dynamics of community structures.
By tuning the value of $\gamma$, we can detect community structure at multiple scales. Specifically, when the value $\gamma$ increases, a network can be divided into more  smaller communities, as shown in Fig.\ref{fig:subfig:1b}.
If we define $\gamma_C$ as the minimal $\gamma$ value for dividing network into $C$ communities, then $\gamma_C$ can be naturally used to indicate the stability or significance of $C$-community structure. For example, Fig 1(c) shows a strong 4-community structure in network $A$ and a weak 4-community structure in network $B$. It can be easily estimated that $\gamma_4(A)<\gamma_4(B)$. Based on the analysis above,the following theorem can be obtained:

$\textbf{Theorem 1.}$ If Hamiltonian function with $\gamma_C$ divide network $G$ into $C$ communities, this result is the weakest one which just meets the definition\cite{Girvan}\cite{Fortunato}, i.e. the number of intra-community edges is equal to the number of inter-community edges.

The proof of theorem 1 is explicit. In addition, the profiles of networks with different scales and types of connectivity can be compared using $\gamma_C$. These differences are defined as ``network distances". For example, Hamiltonian function containing $\gamma_C$ is used to measure the distance between network $m$ and $n$:
\begin{equation} \label{eq:2}
d_{mn}=\Sigma_{\alpha=1}^k|H(\gamma_{\alpha}^{m})-H(\gamma_{\alpha}^{n})|,
\end{equation}
where $H(\gamma_{\alpha}^{m})$ is the Hamiltonian function with parameter $\gamma_{\alpha}^{m}$ in network $m$. ``Network distance'' in this form can be applied without considering the differences of connectivity between various networks, such as size, type of degree distribution and sparsity, and it is convenient to analyze the information hidden behind the topology. However, estimating the value of $\gamma_C$ is a tough job, which can only be tested by optimization methods up to now. In this study, we can use $\gamma_C$ to directly quantify the robustness of a given network. Since few studies have shown the dynamic changes of $\gamma$, we focus on this novel issue and reveal the relationship between $\gamma_C$ and network's robustness in the next section.

\section{ 3. The relationship between $\gamma$ and the robustness}
In this section, a typical case is studied to prove $\gamma_C$ is able to quantify the robustness directly.
For an undirected and unweighted graph $G$ with $N$ nodes and $L$ edges, the topology is characterized by an associated adjacent matrix $A=\{A_{ij}\}$.
$C$ communities are partitioned, and
each community is labeled by $r(r=1,...,C)$.
We denote the number
of inner links connecting each pair of members inside community as $l^r_{in}$, and the number of inter-community links as $l^r_{out}$, i.e. the number of links connecting a member of any one community to a member of another community.
Based on the mentioned notations, there is $L=\sum^C_{r=1}l^r_{in}+\frac{1}{2}\sum^C_{r=1}l^r_{out}$.

Next, the hyper-graph $G^*$ associated to network $G$ is defined as the weighted directed $C$-clique in which each node
corresponds to one community in $G$. In $G^*$, the connection
linking node $r$ to node $s$ is weighted by $\frac{l^{rs}_{out}}{l^r_{in}}$, where $l^{rs}_{out}$ represents the number of links of $G$ that connect members of
the community $r$ with members of the community $s$,
$l^r_{in}$ represents the number of inner links in the source community $r$. The corresponding $C\times C$ Laplacian matrix $\digamma=\{\digamma_{rs}\}$ is
asymmetric, but can be written as a product $\digamma=\Delta\Theta$, where $\Delta=\{\Delta_{rs}\}$ is a symmetric zero row-sum matrix with off-diagonal
elements $\Delta_{rs}=-l^{rs}_{out}$ and diagonal ones $\Delta_{rr}=\sum_{r\neq s}l^{rs}_{out}$ and
$\Theta=diag\{1/l^1_{in},...,1/l^C_{in}\}$. Then $\digamma$ is

\begin{equation}\label{eq:3}
\small
\begin{array}{lcl}
\digamma=\Theta\Delta=\\
\left(\begin{array}{cccc}
\frac{1}{l^1_{in}} & 0 &\ldots& 0\\
0 & \frac{1}{l^2_{in}} &\ldots& 0\\
\vdots & \vdots &\ddots& \vdots\\
0 & 0 &\ldots& \frac{1}{l^C_{in}}\\
\end{array}
\right)
\left(\begin{array}{cccc}
l^1_{out} & -l^{12}_{out} & \ldots & -l^{1C}_{out} \\
-l^{21}_{out} &  l^2_{out} & \ldots & -l^{2C}_{out} \\
\vdots & \vdots & \ddots & \vdots \\
-l^{C1}_{out} & -l^{C2}_{out} & \ldots &  l^C_{out} \\
\end{array}
\right)
\end{array}
\end{equation}
The spectrum of $\digamma$ is non-negative real values, as $\digamma$ is zero row-sum. Then the smallest eigenvalue of $\digamma$,
$\lambda_1^{\ast}$, is zero, while the second smallest one $\lambda_2^{\ast}>0$. The method we proposed to measure the dynamic quality of community structure is defined as follows:
\begin{equation} \label{eq:4}
\xi=\gamma H\lambda_2^{\ast}(\digamma).
\end{equation}
In fact, $H$ is an inherent evaluation function to compute the significance of community structure, based on spectral theory \cite{Our1}. $\lambda_2^{\ast}$ is able to quantify the connectivity
of the hyper-graph, and therefore measure
the extent to which different communities are bounded
and interacted. It should be noticed that both $H$ and $\lambda_2^{\ast}$ are properly normalized, so that even if the
network links were associated to cohesive forces, the two
quantities would be one dimensional.
The maximum of $\xi$ corresponds to a topology in which the community structure is most significant, and thus crucial to this study.

Let us then consider the case that $C$ communities
are all cliques with equal size $N_c=N/C$.
The number of intra-community links $l^r_{in}$, as well as
the number of inter-community links $l^{r}_{out}$ are the same
for all communities.
Then, as $l^r_{in}=l_{in}/C$ and $l^r_{out}=l_{out}/C$, $L=C(l^r_{in}+\frac{1}{2}l^r_{out})=l_{in}+\frac{1}{2}l_{out}$.
Under these assumptions, Hamiltonian function of Eq.(\ref{eq:1}) can be simplified
to the following expression:

\begin{equation} \label{eq:5}
H=C[\frac{1}{L}\frac{(l_{in})}{C}-\gamma(\frac{2L}{C}\frac{1}{2L})^2]=(1-\frac{l_{out}}{2L}-\frac{\gamma}{C}).
\end{equation}
In addition, since $\Delta=\frac{C}{l_{in}}I$, according to the
matrix identity, there is
$\lambda_2^{\ast}=\frac{l_{out}}{l_{in}}$.
Integrating $\lambda_2^{\ast}$ and $\gamma$ with $H$, the function $\xi$ is derived as follows:
\begin{equation} \label{eq:6}
\xi=\gamma(1-\frac{l_{out}}{2L}-\frac{\gamma}{C})\frac{l_{out}}{L-\frac{1}{2}l_{out}}.
\end{equation}

\begin{figure}
\centering
\includegraphics[width=8.5cm,height=5.3cm,angle=0]{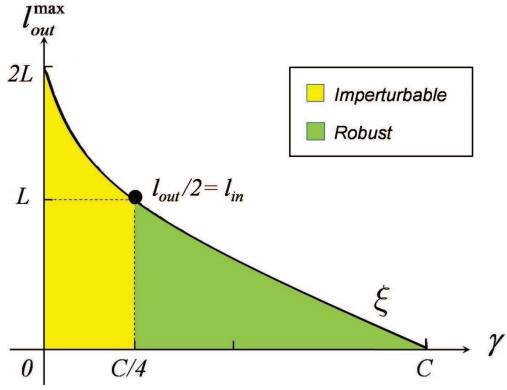} \caption{
The change of $\xi$ with  intra-community edges $l_{in}$. $\xi$ reach the maximal value at $l_{in}^{max}$. At the same time, the community structure is most significant.
}
\label{fig.2}
\end{figure}

To study the dynamic characteristics of networks\cite{Our1}, a particular protocol is adopted by increasing $l_{out}$ at each step from 0 to $2L(1-\frac{\gamma}{C})$, the value at which $\xi$ and $H$ are zero.
Given a fully modularized configuration (in which $l_{out}=0$ and
$l_{in}=L$), we conduct successive rewiring processes, i.e. in
each step an intra-community link corresponding to each community
is deleted, and $C$ inter-community links are formed
by connecting those pairs of nodes (each one in different
communities) which lost their intra-link. In this
way, at the $j$-th rewiring, there are $l_{in}=L-Cj$ and
$l_{out}=2Cj$. Accordingly, the partial derivative of $\xi$, and the dynamical change of inter-community edges $l_{out}$ are calculated as follows:
\begin{equation} \label{eq:7}
\frac{\partial\xi}{\partial l_{out}}=\frac{\gamma L}{(L-\frac{1}{2}l_{out})^2}(1-\frac{\gamma}{C}-\frac{l_{out}}{L}+\frac{l_{out}^2}{4L^2}).
\end{equation}
When $\frac{\partial\xi}{\partial l_{out}}=0$, $l_{out}^{max}=2L(1-\sqrt{\frac{\gamma}{C}})$. The function of $\xi$ reaches the maximum since the second derivative of $l_{out}$ is indeed negative. According to the formation of the maximum of inter-community edges $l_{out}^{max}$, when $\gamma=C$, $l_{out}^{max}=0$. In this case, no inter-community edges exist and the original network cannot be perturbed anymore(increase $l_{out}^{max}$ will decrease $\xi$).
On the contrary, when $\gamma=0$, there is $l_{out}^{max}=2L$. At this time, the network is indeed perturbable because increasing $l_{out}^{max}$ will also increase $\xi$ until all edges are inter-community edges. In this situation, one node is a single community and only belongs to itself.
In a special intermediate case, when $\gamma=\frac{C}{4}$, there is $l_{out}^{max}=L$, and intra-community edges have the same number of edges with inter-community ones. According to the definition\cite{Fortunato}\cite{Girvan}, this is just the threshold testing whether the community structure emerges. As explained above, this $\gamma$ is just the critical value $\gamma_C$, and $\gamma_C$ is closely related to robustness: if the value of $\gamma_C$ increases, the size of imperturbable area(un-robust) is also increases accordingly, as shown in Fig.\ref{fig.2}. This conclusion can be reflected in the following theorem:

$\textbf{Theorem 2.}$ The larger $\gamma_C$,  the lower robustness of a given network including $C$ communities, and vise versa.

For any
given network, robustness information can be derived
by $\gamma_C$ conveniently without particular algorithms, since $\gamma_C$ is only determined by network's topology.

\section{4. A novel method to calculate the critical $\gamma$}

As proved by theorem 2, $\gamma_C$ can be used to quantify a network's robustness. However, the calculation of $\gamma_C$ is a tough task and few studies have addressed this problem. Fortunately, theorem 1 provides a feasible way to the calculation, i.e. $\gamma_C$ can be got through calculating $\gamma$ in a $C$ community structure when $l_{in}=l_{out}$. In this part, a two-stage method is proposed to calculate $\gamma_C$. The detailed procedures are described as follows.

\subsection{4.1. The relationship between $\gamma$ and community edge density}

In unweighted graphs, the ``inverse adjacency matrix" $P$ is defined as $P_{ij}=1$ if there is no edge between node $i$ and $j$ preset, and $P_{ij}=0$, vise versa. In general, we set $P_{ij}=1-A_{ij}$. Since the inner sum of $A_{ij}$, i.e. $\sum_{i\neq j}A_{ij}\delta_{C_i,C_j}$, is the number of edges
in community $r$, and the sum of $P_{ij}=1-A_{ij}$, i.e. $\sum_{i\neq j}(1-A_{ij})\delta_{C_i,C_j}$, is the number of
missing edges in community $r$, Eq.(\ref{eq:1}) can be rewritten as
\begin{equation} \label{eq:8}
H=\sum_{r}(l_{in}^r-\gamma(e^r-l_{in}^r)),
\end{equation}
where the inner sum of $A_{ij}$ has been rewritten in terms of the
number of existing edges $l_{in}^r$, $e^r$ is maximal number of edges that community $r$ would have and the number of missing edges is $e^r-l_{in}^r$.
With minor rearrangement, Eq.(\ref{eq:1}) can be transformed in an ``edge density'' form:

\begin{equation} \label{eq:9}
H=\sum_{r}(e^r(\rho^r-\gamma(1-\rho^r))),
\end{equation}
where the edge density $\rho^r$ of community $r$ is defined as $\rho^r=\frac{l_{in}^r}{e^r}$.

If the energy of a given community is attractive and have a binding pattern, the
term $\rho^r-\gamma(1-\rho^r)$ must be positive.
Rearranging Eq.(\ref{eq:9}) provides a relationship between the resolution parameter $\gamma$ and the critical (minimum) edge density $\rho^*$:
\begin{equation} \label{eq:10}
\rho^r>\frac{\gamma}{1+\gamma}=\rho^*.
\end{equation}
Based on Eq.(\ref{eq:10}), two important inferences can be obtained:
\begin{equation} \label{eq:11}
\rho^*=\frac{\gamma}{1+\gamma},
\end{equation}
and
\begin{equation} \label{eq:12}
\gamma=\frac{\rho^*}{1-\rho^*}.
\end{equation}

\subsection{4.2. The co-evolution model}
We set $\rho^*=\alpha$, and substitute $\gamma=\frac{\alpha}{1-\alpha}$ into Hamiltonian function of Eq.(\ref{eq:1}):
\begin{equation} \label{eq:13}
H=\sum_{i\neq j}(A_{ij}-\frac{\alpha}{1-\alpha}P_{ij})\delta_{C_i,C_j}.
\end{equation}
Then, the term $1-\alpha$ is extracted and an equivalent function is derived
\begin{equation} \label{eq:14}
H^{*}=\sum_{i\neq j}((1-\alpha)A_{ij}-\alpha P_{ij})\delta_{C_i,C_j}.
\end{equation}
In Eq.(\ref{eq:14}), we consider $\alpha$ as a special probability, and the value of $\alpha$ lies between 0 and 1. To optimize Eq.(\ref{eq:14}), the following steps are needed: at each step, a vertex $i$ is picked
randomly. If its degree $k(i)=0$, nothing happens. For $k(i)>0$, (i) with
probability $1-\alpha$, a random neighbor $j$ of $i$ is selected and we put node $j$ into the same community of $i$, i.e. set $\delta_{C_i,C_j}=1$;
(ii) otherwise, with probability $\alpha$, an edge attached to vertex $i$ is selected
and the other end of this edge is rewired to a randomly chosen vertex in the same community with $i$.
This process continues until no edge connecting individuals between different communities.

This dynamical evolutionary process
can be considered as a special case of famous Holme-Newman model\cite{Holme}. There are two extremes corresponding to the value of $\alpha$.
When $\alpha=1$, only rewiring steps(step ii) occur. Once all of $L$
edges are touched, the graph has been split into
$C$ components, each consisting of individuals who share the same
label. Because none of the states have changed, the components
are small (i.e., their sizes are Poisson distribution with mean $\frac{N}{C}$).
According to classical results for the coupon collector's problem\cite{Durrett}, $L\log L$ updates are approximately required. In contrast, for $\alpha=0$, this system reduces to the voter model
on a static graph. If we suppose that the initial graph is an ER random graph in which each vertex has average degree
$\langle k\rangle > 1$, and then there is a "giant component"
that contains a positive fraction $\langle k\rangle N$ of the vertices, and
the second largest component is small having only $O\log(N)$ vertices. The voter model on the
giant component will reach a consensus in $O(N^2)$ steps.

\begin{figure}
\centering
\includegraphics[width=8.5cm,height=5.5cm,angle=0]{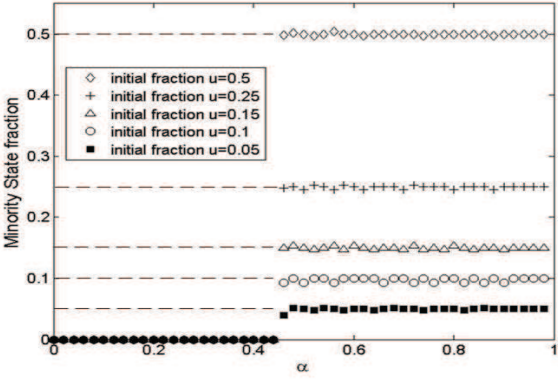} \caption{
The change of fraction in minority state with $\alpha$ on ER random graphs with $N = 10000$ nodes and average degree $k=4$. The fraction in minority state $u$ equals to $0.5,0.25,0.1$ and $0.05$, respectively.
}
\label{fig.3}
\end{figure}

To pursue the study, a two state Potts model (the two different spin states called 0
and 1) is proposed instead of a number proportional to the size of the graph. This model is also called Ising model. As the same as Holme-Newman model, the final fraction $u$ of nodes with the minority spin states undergoes a discontinuous transition at a value $\alpha$ that does not depend on the initial density. Fig.\ref{fig.3} shows results of simulations for our method starting from an initial
graph that is ER random graph with $N=10000$ nodes and average degree $k=4$. Spin values are initially assigned randomly with the probability of state 1 given by fraction $u=0.5, 0.25, 0.1$, and $0.05$. The figure shows the final fraction $u$ of nodes with the minority spin states from five scenarios for each $u$. Although the fraction of nodes with state 1 is less than 0, this minority state will reach a stable state instead of being "assimilated" or "swallowed" by state 0. In community analysis, this phenomenon is equivalent to free of "resolution limit" problem  pointed out by Fortunato et al\cite{Fortunato}, where the modularity $Q$ exists at an intrinsic scale beyond which small qualified communities cannot be detected by maximizing the modularity.

\subsection{4.3. The computation of $\alpha_C$}
Since $\gamma=\frac{\alpha}{1-\alpha}$ analyzed above, $\alpha$ is proportional to the value of $\gamma$. If $\alpha_C$ is defined as the critical threshold of $\alpha$ when $l_{in}=l_{out}$ in $C$ community structure, and then estimate that the value of $\gamma_C$ is proportional to the critical threshold $\alpha_C$, using the relationship $\gamma_C=\frac{\alpha_C}{1-\alpha_C}$. Hereby, we focus on the case $C=2$ and find the solution which can be extended to more general cases.

For methodology, we use mean-field theory with Markov dynamical process to compute $\alpha_C$. First, Let $L_{xy}$ be the
number of edges of adjacent nodes with states $x$ and $y$, and $L_{xyz}$ is defined as the
number of oriented triples $x-y-z$ of adjacent sites with states $x, y,
z$, respectively. $L_{01}$ is equal to $L_{10}$, for example, in the 0-1-0 case, all such triples will be counted
twice, but the approach is limited of dense graphs, where the general statistics are the
number of homomorphisms of some small graphs (labeled by ones
and zeros in our case) into the random graph being studied.
It is common to use the pair approximation (PA), which in essence assumes that
the equilibrium state is a Markov chain: $L_{100}=L_{10}L_{00}/\omega N_0$, where $N_0$ is the number of vertices in state 0, and $\omega$ is the clustering coefficient of a network. Using mean field theory and algebraic transformation, the following theorem can be deduced:

$\textbf{Theorem 3.}$ Defining $\langle k\rangle$ as the average degree, $u$ as the fraction of nodes with minority spin state, $N$ as the number of nodes, $\omega$ as the clustering coefficient, then, the critical threshold $\alpha_C$ and the number of inter-community edges satisfies
$\alpha_C=1-\frac{\omega}{\langle k\rangle}$ and $L_{01}=Nu(1-u)(\langle k\rangle-\frac{\omega}{1-\alpha})$, respectively.

\textbf{Proof.} The calculations presented here are inspired by similar equations in Kimura and
Hayakawa\cite{Kimura}. According to the mechanism of our model, by
considering all of the possible changes, the partial differential equations can be established as follows:

\begin{equation} \label{eq:15-1}
\frac{1}{2}\frac{\partial L_{10}}{\partial t}=-2L_{10}+(1-\alpha)[L_{011}-L_{101}+L_{100}-L_{010}],
\end{equation}

\begin{equation} \label{eq:15}
\frac{1}{2}\frac{\partial L_{11}}{\partial t}=L_{10}+(1-\alpha)[L_{101}-L_{011}],
\end{equation}
and
\begin{equation} \label{eq:16}
\frac{1}{2}\frac{\partial L_{00}}{\partial t}=L_{10}+(1-\alpha)[L_{010}-L_{100}].
\end{equation}
The fact that this notation is more natural than dividing by 2 to eliminate overcounting, can be seen by
observing that, if $k(x)$ is the degree of node $x$, there are $\sum_{ij}L_{ij}=\sum_{x}k(x)$ and $\sum_{ijk}L_{ijk}=\sum_{x}k(x)[k(x)-1]$. Also, $L_{11}+2L_{10}+L_{00}=L$, where $L$ is the number of edges, and the sum of the three differential
equations(i.e. Eq.(\ref{eq:15-1})-Eq.(\ref{eq:16})) is zero.

Taking steady solution of these equations and the pair approximation as before, we get
\begin{equation} \label{eq:17}
\frac{1}{1-\alpha}L_{10}=\frac{L_{01}L_{11}}{\omega uN}-\frac{L_{10}L_{01}}{\omega(1-u)N},
\end{equation}
and
\begin{equation} \label{eq:18}
\frac{1}{1-\alpha}L_{10}=\frac{L_{10}L_{00}}{\omega(1-u)N}-\frac{L_{01}L_{10}}{\omega uN}.
\end{equation}
Eq.(\ref{eq:17}) and Eq.(\ref{eq:18}) lead to the equations
\begin{equation} \label{eq:19}
\frac{L_{11}}{\omega uN}-\frac{L_{10}}{\omega(1-u)N}=\frac{1}{1-\alpha},
\end{equation}
and
\begin{equation} \label{eq:20}
\frac{L_{00}}{\omega(1-u)N}-\frac{L_{10}}{\omega uN}=\frac{1}{1-\alpha}.
\end{equation}

Adding $uN$ times Eq.(\ref{eq:19}) to $(1-u)N$ times Eq.(\ref{eq:20}), we have
\begin{equation} \label{eq:21}
L_{11}+L_{00}-(\frac{u}{1-u}+\frac{1-u}{u})L_{01}=\frac{\omega N}{1-\alpha}.
\end{equation}
When $L_{01}=0$, only intra-community edges exist, we have $L_{11}+L_{00}=\langle k\rangle N$. The threshold information can be got:
\begin{equation} \label{eq:21}
\alpha_C=1-\frac{\omega}{\langle k\rangle}.
\end{equation}
Using Eq.(\ref{eq:21}) and $L_{11}+L_{00}=\langle k\rangle N-2L_{01}$, we have

\begin{equation} \label{eq:22}
[\langle k\rangle-\frac{\omega}{1-\alpha}]N=\frac{L_{01}}{u(1-u)}.
\end{equation}
The number of inter-community edges satisfies
\begin{equation} \label{eq:23}
\frac{L_{01}}{N}=u(1-u)(\langle k\rangle-\frac{\omega}{1-\alpha}).
\end{equation}

The proof is completed.

The approach is limited to dense graphs. As $\gamma_C=\frac{\alpha_C}{1-\alpha_C}$, $\alpha$ is proportional to the value of  $\gamma$, and then we can get $\gamma_C=\frac{\langle k\rangle}{\omega}-1$. This approximation is simple and convenient to compute in large scale networks, by using sampling technology. Although $\gamma_C$ is derived from two states Ising model, it can be directly applied to network with more than 2 communities, since the elements $\langle k\rangle$ and $\omega$ are only determined by network topology without using any partition algorithm.

\section{ 5. Experiments }

We test the index on both the classical GN benchmark presented by Girven and
Newman\cite{Girvan} and the more challenging
LRF benchmark proposed by Lancichinetti, Fortunato and
Radicchi\cite{Lancichinetti01}. GN network has $n = 128$ nodes that are
divided into 4 communities with 32 nodes each. Each
node is connected to average $\langle k^{in}\rangle=16$ nodes of its own
group and $\langle k^{out}\rangle$ of the rest of the network. The total
degree of each node is always kept constant and equals
to $k=\langle k^{in}\rangle+\langle k^{out}\rangle$.
In the LFR benchmark, each node is
given a degree taken from a power law distribution with
an positive exponent.
Additional, each node shares a fraction $\theta$ with
other nodes in the network, where $\theta$ is the mixing parameter.
The clarity of community structure can be adjusted
by the mixing parameter $\theta$.

\begin{figure}
\center
  \subfigure[]{
    \label{fig:subfig:4a} 
    \setcounter{subfigure}{1}
    \includegraphics[width=4.1cm,height=3.35cm]{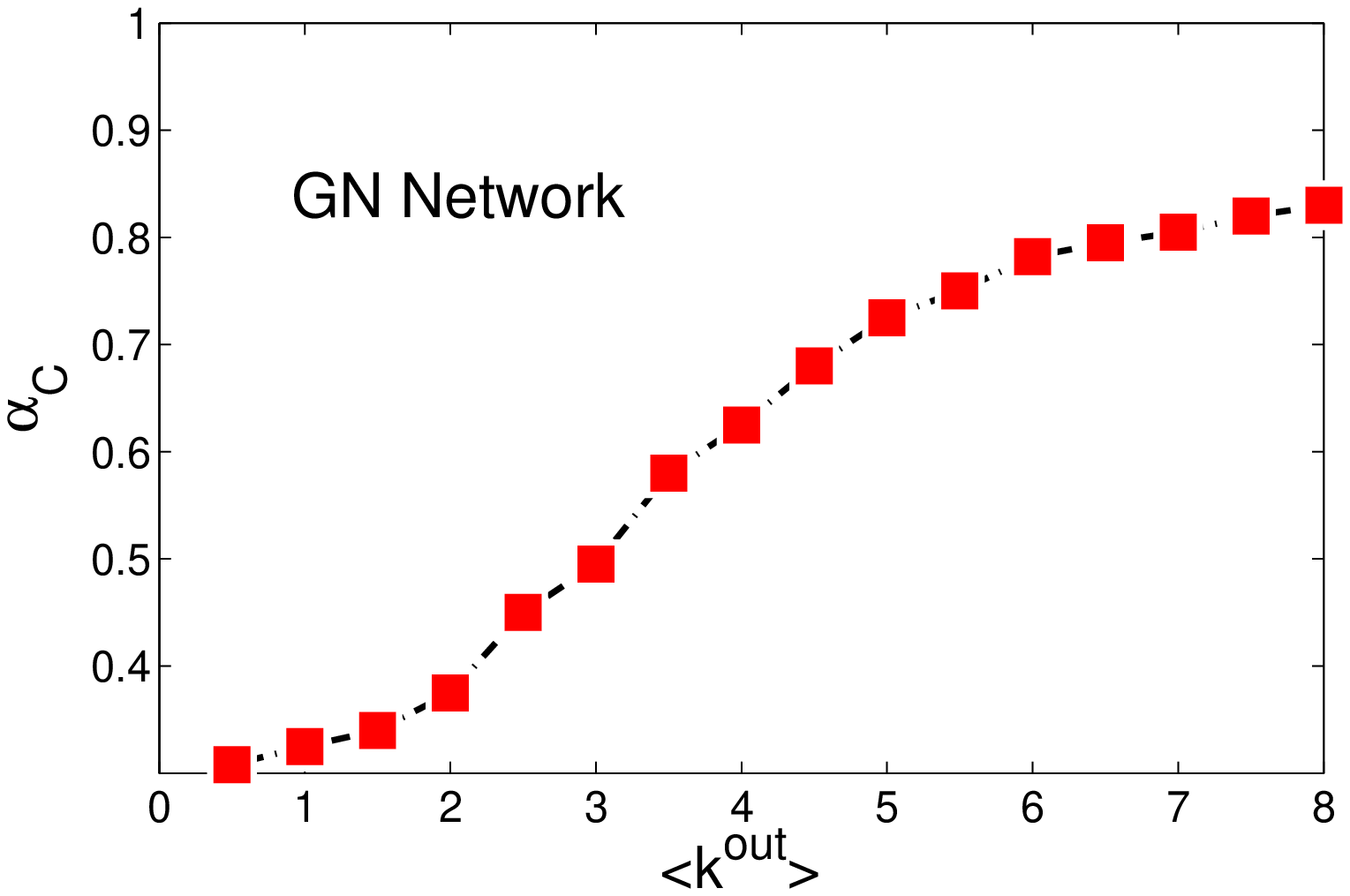}}
  \subfigure[]{
    \label{fig:subfig:4b} 
    \setcounter{subfigure}{2}
    \includegraphics[width=4.1cm,height=3.35cm]{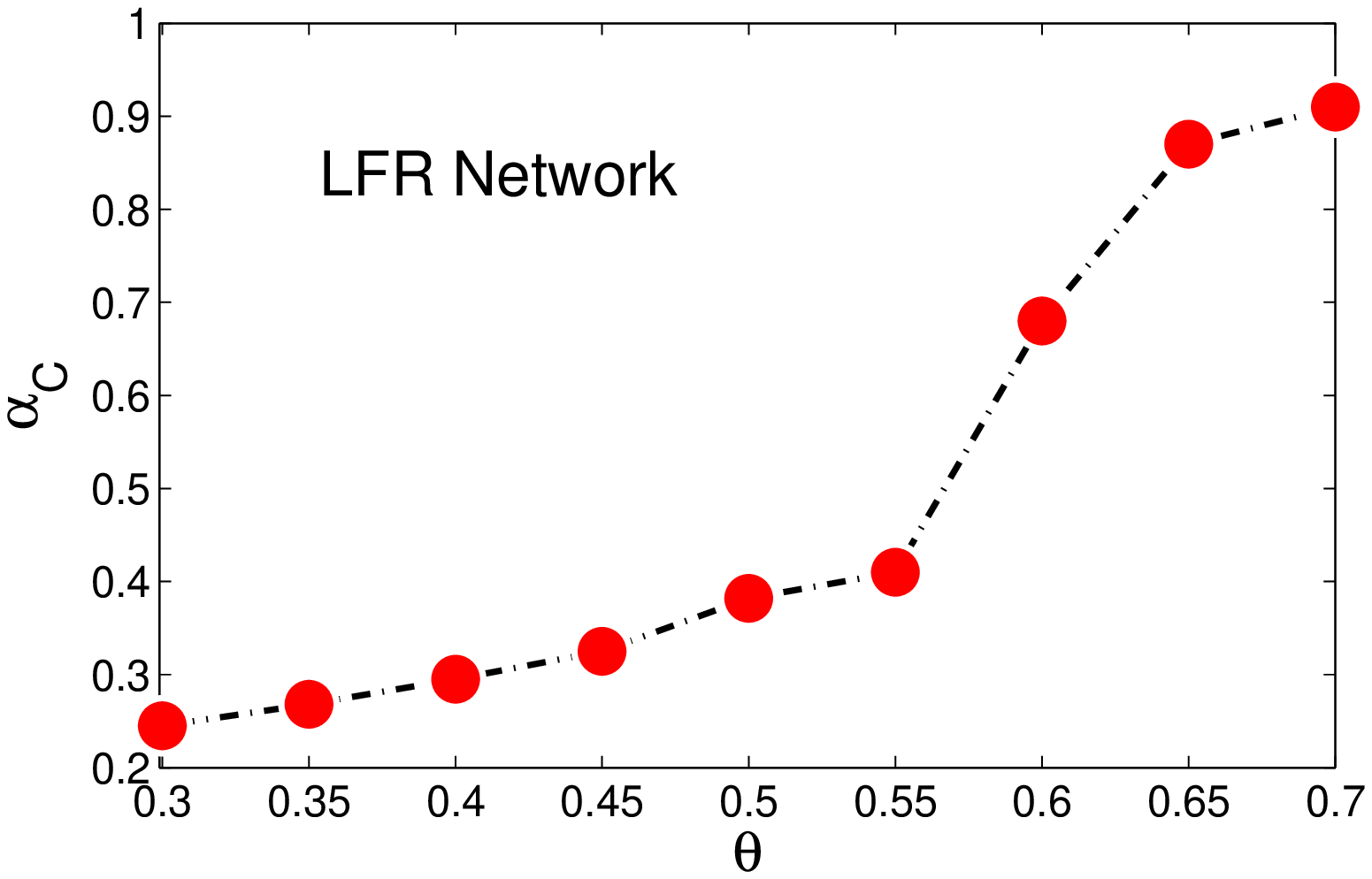}}
\caption{The performance of $\alpha_C$ on both GN and LFR network. (a) In GN network, $\alpha_C$ increases with the increase of $k^{out}$. The community structure varies from clear to vague in accordance to $\alpha_C$ value from 0.3 to 1. (b)In
LFR-benchmark, the average degree $k=20$, and maximum degree
is 50 and $P(k)\propto k^\gamma$. Maximum and minimum community
sizes are 50 and 20 respectively. With the
increase of mix parameter $\theta$, the $\alpha_C$ index increases.} \label{fig.4}
\end{figure}

As is well known, the communities
become fuzzier and thus more difficult to be identified
when $\langle k^{out}\rangle$ and $\theta$ increase. Hence, the robustness of the community structure will tend to be weaker and the $\alpha_C$
index will increase. The numerical results of $\alpha_C$ value for both $\langle k^{out}\rangle$ and $\theta$
are shown in Fig.\ref{fig.4}. The figure indicates that the index $\alpha_C$ works
well in these networks: when community structure
is very clear, the $\alpha_C$ is near
0.2-0.3; when the network
is nearly a random one, the corresponding $\alpha_C$ is very close to 1. Thus, this method shows a great ability in characterizing the properties of modular structure and the lower the $\alpha_C$(or $\gamma_C$) index is, the
more robust community structure will be.

In order to verify our method, it is also applied to three famous artificial networks-- ER random graph, BA scale-free network, and $P\&S$ network\cite{Papadopoulos} where the number of nodes are 10,000 and the average degree is all 3. The experimental result indicates that $P\&S$ model is the most robust one. We also test the method on real networks and the results are shown and analyzed in Supplementary Material\cite{Appendix}.

\section{ 6. Conclusion }

In summary, this letter presents a new community analysis method which is able to uncover the connection between robustness of community structure and the critical threshold of resolution parameter $\gamma_C$. Based on the theoretical analysis, a novel computation method is developed to quantify $\gamma_C$ using co-evolution theory. The effectiveness and efficiency are demonstrated and verified, which can be applied to broad problems in network analysis and data mining due to its solid mathematical basis and efficiency.

\acknowledgments
We are grateful to the anonymous reviewers for their valuable suggestions. The authors are separately supported by NSFC grants 71401194, 71403304, 11131009, 91324203 and ``121'' Youth Development Fund of CUFE grants QBJ1410.

\end{document}